\documentclass{aastex}
\usepackage{spr-astr-addons}
\usepackage{url}

\RequirePackage{color}

\newcommand{\emaila}{kunjaya@as.itb.ac.id}

\begin{document}

\title{Can Self Organized Critical Accretion Disks Generate a Log-normal Emission Variability in AGN?}
\slugcomment{}
\shorttitle{SOC Generate Lognormal Emission}
\shortauthors{Kunjaya et al.}

\author{C. Kunjaya\altaffilmark{1,2}}
\author{P. Mahasena\altaffilmark{1}}
\author{K. Vierdayanti\altaffilmark{1}} \and \author{S. Herlie\altaffilmark{1}}
\altaffiltext{1}{Dept. of Astronomy, Institut Teknologi Bandung, Jl. Ganesa 10,
Bandung 40132, Indonesia \\ e-mail : \emaila}
\altaffiltext{2}{Bosscha Observatory, Lembang, Indonesia}

\begin{abstract}
Active Galactic Nuclei (AGN), such as Seyfert galaxies, quasars, etc., show light variations in all wavelength bands, with various amplitude and in many time scales. The variations usually look erratic, not periodic nor purely random. Many of these objects also show lognormal flux distribution and RMS - flux relation and power law frequency distribution. So far, the lognormal flux distribution of black hole objects is only observational facts without  satisfactory explanation about the physical mechanism producing such distribution in the accretion disk. One of the most promising models based on cellular automaton mechanism has been successful in reproducing PSD (Power Spectral Density) of the observed objects but could not reproduce lognormal flux distribution. Such distribution requires the existence of underlying multiplicative process while the existing SOC models are based on additive processes. A modified SOC model based on  cellular automaton mechanism for producing lognormal flux distribution is presented in this paper. The idea is that the energy released in the avalanche and diffusion in the accretion disk is not entirely emitted instantaneously as in the original cellular automaton model. Some part of the energy is kept in the disk and thus increase its energy content so that the next avalanche will be in higher energy condition and will release more energy. The later an avalanche occurs, the more amount of energy is emitted to the observers. This can provide multiplicative effects to the flux and produces lognormal flux distribution.\end{abstract}


\section{Introduction}
Black hole objects, in a wide range of mass from  Active Galactic Nuclei (AGNs)
to black hole X-ray binaries (BHBs), generally show light fluctuation which is
not periodic, nor purely random.
This kind of variability is found in all wavelengths and in a wide range of
time scales. It is one of the keys for understanding the nature of these
objects. 

Long term monitoring of AGN light curves still fail to define clearly
the upper limit of variability time scale, constraining our ability to extract 
information about disk structure from variability studies \citep{hawkins06}. 
For instance, twenty eight years of monitoring program of about 1500 quasars by \cite{hawkins96} 
still did not reach the common upper limit of quasars variability time scale. However, analysis
of quasars long term light curve in quasars rest frame using Fourier power spectrum yield a break 
frequency. \cite{hawkins07} interpreted it as the 11 year characteristic time scale of quasars.

Several attempts have been done to reproduce such variability, for example, \cite{Terrell72}
introduced a shot noise model, which was developed further by several authors
\citep[e.g.][]{{Lehto89, Lochner91}}. 
This model could reproduce the observed light curve and power spectral density
(PSD) of the AGNs by superposition of the shots that occur randomly. 
In addition, the shots have identical time profiles with exponential decays.
However, there is no clear physical explanation for the creation of the shots \citep{Takeuchi97}.
Other models of variability that have been proposed so far, include starburst, disk instability 
and microlensing models \citep[][and references therein]{hawkins04}.

\cite{mineshige94a, mineshige94b}
have tried to simulate the fluctuation in the light curve of black hole
objects under the notion of self organized criticality (SOC) by using a cellular
automaton rule. SOC is a continuous critical state reached spontaneously by a system receiving 
energy and/or matter from outside, independent of its initial condition.

The concept of SOC was first introduced by Bak et al. (1998),
which used sand pile system as 
an illustration. When grains of sand are dropped continuously in a surface, it will accumulate and create 
a peak which is higher and higher. When the peak is high enough and the slope of some part of 
the pile reach a critical value, then an additional falling grain of sand will trigger another grain 
to fall one after the other and creating an avalanche. After an avalanche the slope of the pile 
will be shallower. Additional sand drop will make the surface of the pile steeper and steeper, 
finally reach a critical value and trigger another avalanche, and so on. 
The sand pile will always be close to the critical state which may trigger avalanches. 
If the energy release of this sand pile system is recorded in a time series, it will produce $1/f$-like
variability in its PSD.

In their model, \cite{mineshige94a} and Mineshige et al. (1994b)
divided the two-dimensional disk in radial and azimuthal
directions, turned into cells.
The disk receives mass input from a randomly chosen outermost cell and finally loses
mass through the innermost ring due to accretion onto its central black hole. 
Similar to the sand pile model, in the accretion disk the mass is transfered inward through 
avalanche and diffusion processes that may happen in any cell under some special conditions.
These processes convert gravitational potential energy into heat, eventually
radiated from the disk.
By using this model, they could successfully simulate the light fluctuation
and its PSD. The PSD showed a $1/f$ tendency with power law
index similar to those observed from some black hole objects.

However, later observations of some AGNs revealed that some of them
show lognormal X-ray flux distribution \citep[e.g.][]{Gaskell04} and follow 
linear RMS-flux relation \citep[e.g.][]{uttley05}. In this case RMS is the 
root mean square of the flux and the flux means the average flux for the 
chosen bin. These properties are also found in some galactic accreting source, such as
Cyg X-1 and SAX J1808.4-3658 \citep{uttley01, Gleis04}.
The SOC model introduced by \cite{mineshige94a, mineshige94b} could not explain such
phenomena \citep[see e.g.][]{Gaskell04}, because, multiplicative processes are
necessary to yield lognormal flux distribution, whereas the underlying process
in the original SOC model is additive. 
What kind of physical processes in the accretion disk can yield lognormal
flux distribution?
Some mathematical models have been introduced \citep[see e.g.][]{Gaskell04, uttley05}, but without a clear explanation on the physical mechanisms.

In this work, we introduce an idea with reasonable physical mechanism which has multiplicative effect, in an attempt to obtain a lognormal flux distribution.
We assume that the accretion flow is dominated by advective cooling instead of
radiative cooling. This mechanism was called advection-dominated accretion flow (ADAF) \citep{Ichimaru77, Abra88, Kato08}.

In the present work, we adopt the idea that the energy released by the viscous
process is not efficiently emitted locally which could act as a multiplicative
effect that yield lognormal flux distribution.
It is important to note that \cite{Takeuchi97} created a numerical
simulation of hydrodynamical model of advection dominated accretion disk with
critical behavior. They aimed to reproduce $1/f$-like X-ray fluctuation with
a physical mechanism focusing on the Cyg X-1 type fluctuation.
We, on the other hand, aimed to address the lognormal flux distribution problem
that was claimed to be unable to be reproduced by a cellular automaton model
\citep[see e.g.][]{Gaskell04}.

The plan of this paper is as follows:
We will describe the modified cellular automaton model in the second section in which
we will also briefly review the lognormal distribution and the original
cellular automaton model.
In section three, we present the simulation details. 
The results and analysis will be presented in section 4.
Section five is devoted for discussion and conclusion.

\section{The Model}

\subsection{Lognormal Distribution}
The original SOC model, due to its additive nature, will have a normal
flux distribution.
The normal distribution can be describe by the  following formula:
\begin{equation}
\Phi (x) = \frac{1}{\sigma \sqrt{2 \pi}} \exp \left[ -\frac{1}{2} \left( \frac{x-\mu}{\sigma} \right)^{2} \right],
\end{equation}
where $\mu$ and $\sigma^{2}$ are the mean of the distribution and the variance
of the distribution, respectively.

A lognormal distribution is similar to a normal distribution, but the
variable is replaced by a logarithmic form
\begin{equation}
\Phi (x) = \frac{1}{\sigma \sqrt{2 \pi}(x-\tau)} \exp \left[ -\frac{1}{2} \left( \frac{\log(x-\tau)-\mu}{\sigma} \right)^{2} \right],
\end{equation}
where $\tau$ is the threshold parameter representing the lower limit on $x$.

The lognormal distributions, which can describe many phenomena such as economic growth,
population statistics, grain size in sand, etc., are also found in the X-ray flux
distribution of some black hole objects \citep[e.g.][]{Gaskell04}. The fundamental
difference between both distributions is their underlying process,
additive in normal and multiplicative in lognormal distribution.

\subsection{Advection Process}
In this work, in order to create multiplicative mechanism in accretion disks of AGNs,
we adopt the idea of the existing theory of advection process in accretion disk, 
which is well known as ADAF.

Theoretically, there are two regimes of ADAF \citep[see][for details]{Kato08}. 
The first regime is better known as radiatively inefficient
accretion flow (RIAF) in which the radiation energy from the viscous process is not
efficiently emitted locally (as in the case of the standard model) but stored
as gas entropy and transported inward. Due to inefficient radiative cooling, 
the gas becomes so hot that high-energy
emission, in the form of hard X-ray, can be produced.
This regime is also associated with a very low mass accretion rate.
Such model was first developed by \cite{Ichimaru77} and extensively discussed,
later, by \cite{Narayan94, Narayan95a, Narayan95b}, \cite{Abra95, Abra96} and others. 
\cite{Narayan95c} showed that such model can reproduce the spectra of Sgr
A$^{\ast}$. It can also reproduce the spectra of some black hole binaries in
their low luminosity state (also known as low/hard state).     

The other regime of ADAF occurs when the mass accretion rate is moderately
high, close to or even exceeding the critical mass accretion rate,
$\dot{M}_{crit} \equiv L_{\rm E}/c^2$ and thus also known as supercritical accretion
flow. The model of this ADAF regime was first developed by \cite{Abra88} 
and known as the slim disk model.
Unlike RIAF, in slim disk model the advective cooling does not always exceed
the radiative cooling, especially when $\dot{M}$ is barely exceed the critical
value \citep{Abra88}. 
The important key for understanding supercritical accretion is the photon
trapping effect. At high mass accretion rate, the disk becomes so optically
thick that large number of interactions between photons and matter delay the radiation
of energy released by the viscous process, especially from the part near the
central region of the disk \citep[see][]{Ohshuga03, Ohshuga05}.
That is, the radiation energy is trapped in the material flow and advected
inward. Some of this trapped radiation energy is released at smaller
radii and some fall onto the black hole with the accreting material without
being radiated away.
Supercritical accretion has been observed in Galactic microquasars, e.g. GRS1915+105. 
Some ultraluminous X-ray sources (ULXs), and narrow-line Seyfert 1 galaxies
(NLS1s) also seem to exhibit supercritical accretion \citep[see e.g.][]{Watarai01, Kawa03}.

\subsection{Modified SOC Model}
In the original model 
\citep{mineshige94a, mineshige94b} an input of mass $m$ is supplied at a
randomly chosen cell in the outermost ring. From each ring there
is mass diffusion $m'$ from a randomly chosen cell to its adjacent
inner cell. Suppose that $i$ is index in radial and $j$ in azimuthal direction, 
if $M_{i,j}$ is the mass of cell $(i,j)$ where the diffusion occur, 
the mass will change according to the following rule :
\begin{eqnarray}
M_{i,j}\rightarrow M_{i,j}-m' \nonumber \\ M_{i+1,j}\rightarrow M_{i+1,j}+m'
\end{eqnarray}
Next, the model search the unstable sites which satisfy the instability
criterion $M_{i,j}>M_{crit}(r)$, where $M_{crit}(r)$ is a given
critical mass. If,  $M_{i,j}>M_{crit}$ then avalanche will occur
according to the following rule :
\begin{eqnarray}
M_{i,j}\rightarrow M_{i,j}-3m \nonumber \\
M_{i+1,j-1}\rightarrow M_{i+1,j-1}+m \\
M_{i+1,j}\rightarrow M_{i+1,j}+m \nonumber \\
M_{i+1,j+1}\rightarrow M_{i+1,j+1}+m \nonumber
\end{eqnarray}

The diffusion and the avalanche processes will release gravitational
energy and produce radiation whose luminosity proportional
to the loss of gravitational potential energy. \cite{Kunjaya05} made a 
modification of the cellular automaton rule to incorporate mass input not only 
via a cell but also all of the cells in the outermost ring receive small amount of
matter from outside. 

The aim of the modification is to search for a mechanism which 
can agree better to AGN observational facts. Unlike the case for the binary stellar 
mass black hole which receive mass input from the companion through Lagrangian point only, 
it is more logical to think that mass input to the AGN's accretion disk comes from many 
sources surrounding the disk.

The original SOC model could not yield lognormal flux distribution because of
its intrinsic additive processes through avalanche and diffusion. 
For each avalanche or diffusion process, the energy
released is 
\begin{equation}
L \propto \sum_i\left(\frac{1}{r_{i+1}}-\frac{1}{r_i}\right) \Delta m_i,
\end{equation}
where $\Delta m_i$ is the total amount of mass transported from
the ring $i$ to $i+1$ in a time step by avalanche and diffusion. 
The energy produced from an avalanche in a cell is simply added to those of the
other cells in the same time step, and assumed that the energy release is
totally radiated locally. 

It is reasonable to think that not all of the energy produced are radiated
locally. In other words, some fraction of energy remains in the disk
and increases the energy content of the disk.
Such mechanism has been considered in ADAF model 
\citep[e.g.][]{Abra88}. However it was not applied in the framework of
SOC model.

In the other avalanche, the part of the stored energy will also be radiated
along with the currently produced energy. This will create a multiplicative
effect needed to yield a lognormal flux distribution. 
So, when a succeeding avalanche occurred in the same time step, the energy emitted
is larger because the energy content of the disk has increased due to the
absorption of some energy from the previous avalanche. 

Assume that the energy produced from an avalanche or diffusion is
\begin{equation}
E_j = \left(\frac{1}{r_{i+1}}-\frac{1}{r_i}\right) \Delta m_j.
\end{equation}
Let the fraction of the energy emitted from an avalanche or diffusion is $a$,
then the emitted energy is 
\begin{equation}
L_1 = aE_1.
\end{equation}
The rest, $S_j$ will be stored in the disk,
\begin{equation}
S_1 = (1-a)E_1.
\end{equation}
In the next avalanche, still in the same time step, the energy produced is $E_2$,
a part of it will be emitted,
also a part of the stored energy from the previous avalanche will be emitted:
\begin{equation}
L_2 = L_1 + aE_2 + aS_1.
\end{equation}
The energy stored in the disk will be:
\begin{equation}
S_2 = (1-a)S_1 + (1-a)E_2 = (1-a)(S_1 + E_2).
\end{equation}
This energy will influence the amount of energy emitted in the next avalanche.
Then, in general, the energy emitted after the $n$-th avalanche, will be :
\begin{equation}
L_n = L_{n-1} + a(S_{n-1} + E_n).
\end{equation}
And the stored energy will be:
\begin{equation}
S_n = (1-a)(S_{n-1} + E_n).
\end{equation}

A black hole has a very strong gravity, even light can not escape from its gravitational pull, 
so that eventually there will be a certain amount of energy produced in the 
avalanches which will be swallowed by the black hole. In this model it is assumed 
that the amount of energy that falls into the black hole is equal to the amount 
of energy stored in the disk after the last avalanche in the same time step. 
Such mechanism occur in the slim disk model, in which the
energy released by avalanche, diffusion or viscous flow is stored as radiation
entropy and transported inward with accretion \citep{Kato08}.
The balancing between the remaining energy in the disk and the energy
absorption by black hole may sound artificial, but it is necessary to keep
energy balance in the disk so that the disk always in a stable critical state.

\section{Simulation}
We did a numerical simulation based on cellular automaton mechanism which was previously done by \cite{Kunjaya05}. The main aim of the simulation is to test whether it is possible to generate lognormal flux distribution. Therefore we chose a set of fixed values of the previous parameters, that is $m$, the amount of each avalanche $m'$, the amount of diffusion and $m''$, the amount of random mass input in the outermost ring of the accretion disk. 

Each time the avalanche or diffusion occurs, the amount of energy produced are calculated using equation (5). The accumulated energy to be emitted after a series of n avalanche in a time step is calculated using equation (11) and the accumulated energy stored in the disk is calculated using equation (12). 

In the simulation, we tried several values of $a$, $0.01 \leq a \leq 0.8$. The system with $a<0.5$ means most of the dissipated energy from each avalanche or diffusion are not directly emitted but is stored in the disk and to be emitted part by part in the avalanches or diffusions afterwards at smaller radii. This setting is similar to those of advection dominated accretion flow model \citep{Kato08}.    

\begin{figure}[t]
\includegraphics[width=60mm, height=80mm, angle=270]{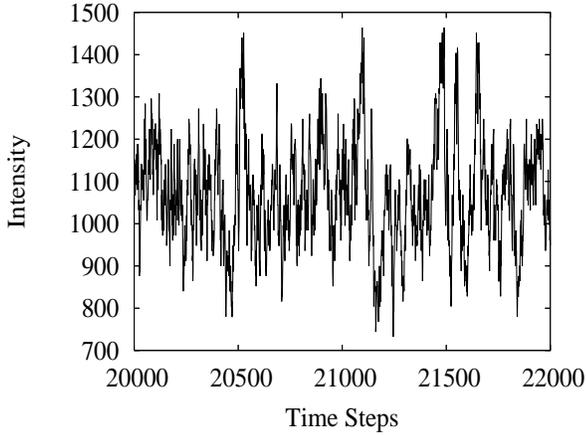}
\caption{Sample of the light curve with $a=0.1$}
\label{fig1}
\end{figure}

We ran the simulation for $10^5$ iterations. The first few thousands calculations were not used in the analysis to avoid the possibility that the system has not reach the SOC state. The sample of the simulated light curve is shown in figure 1. It is roughly similar to the typical light variation of black hole objects, that is irregular, not periodic nor purely random. We then analyzed light curve resulted from the simulation to find the clues of its flux distribution.

\section{Result}

First we checked whether the simulated light curve have power law frequency distribution by calculating its PSD. Figure 2 shows an example of PSD for $a=0.1$ which has power law distribution, and is in good agreement with what were commonly derived from the observed light curve of AGNs. The PSD does not change much with the variation of $a$, and all yield similar slope.

\begin{figure}[t]
\includegraphics[width=60mm, height=80mm, angle=270]{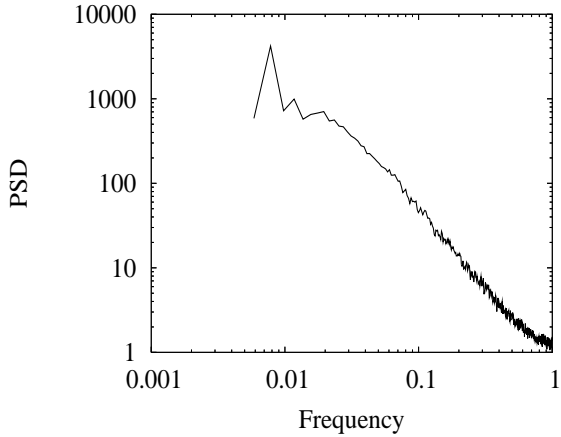}
\caption{Power Spectral Density of the simulated light curve with $a=0.1$}
\label{fig2}
\end{figure}

Another tools which can be used for checking the conformity of this model with observation is structure function \citep{simonetti85}. So far, structure function has been one of the best analysis tool for AGN variability \citep{hawkins04} by which we can derive some information from models and compare to those from the observerd light curve even for under sampled data and uneven sampling. Figure 3 shows the structure function derived from the simulated light curve. It shows a similar characteristics with those generally derived from the observations of AGNs.

\begin{figure}[t]
\includegraphics[width=60mm, height=80mm, angle=270]{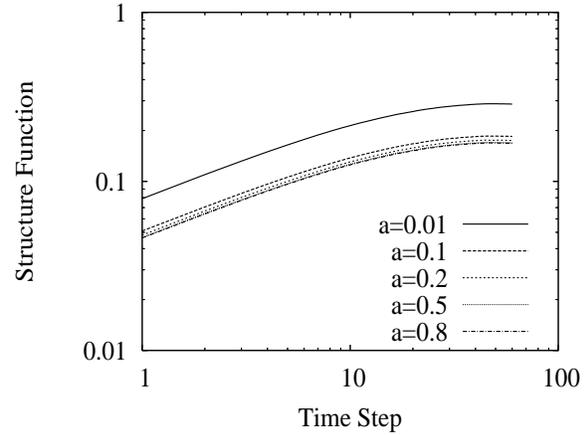}
\caption{Structure functions of the simulated light curve for several values of $a$}
\label{fig3}
\end{figure}

To analyze the flux distribution of the simulated light curve, we grouped the flux into predefined bins and count the number of data in each bin. The width of the bin must be carefully determined, since if the bin is too wide then it will make the number of points in the distribution graph becomes too small while too narrow bins will make the wing of the distribution under sampled.

\begin{figure}[t]
\includegraphics[width=60mm, height=80mm, angle=270]{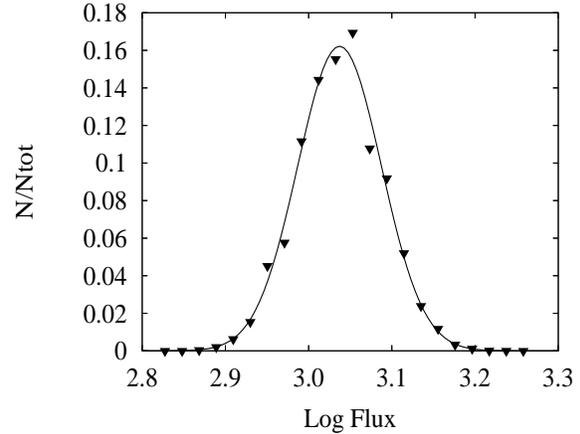}
\caption{Flux distribution for advection domination with $a=0.1$, solid line is the theoretical lognormal function from equation 2 while the solid triangle points are from simulation}
\label{fig4}
\end{figure}

\begin{figure}[t]
\includegraphics[width=60mm, height=80mm, angle=270]{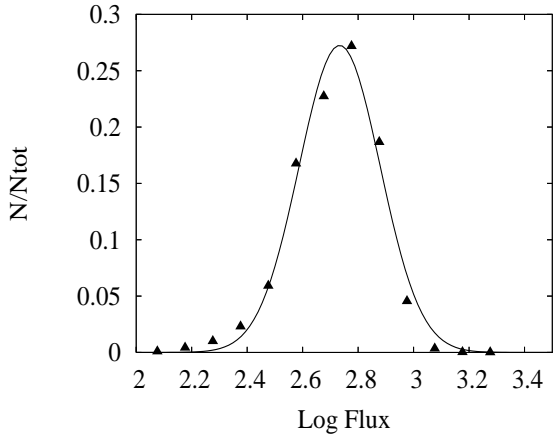}
\caption{Flux distribution for original SOC model (no advection) with $a=1$, solid line is the theoretical lognormal function from equation 2 while the solid triangle points are from simulation}
\label{fig5}
\end{figure}

The flux distribution is then fitted to the lognormal function described in equation (2) 
by adjusting the parameters $\tau$, $\sigma$ and $\mu$. The flux distribution could fit 
well with the model of lognormal distribution with certain set of parameters and is shown in figure 4. 

For comparison, we present the flux distribution of the original
SOC model without ADAF in figure 5. Roughly, it may also look fit, but when we study it more carefully, 
the points position relative to the line, then it is evident that the points in the wing are slightly 
shifted to the left, while near the peak, the points shifted to right relative to the line. 
This means the points will be better fitted to a skewed distribution. 
The skewness of the other case with $0.1 < a < 1$ is less clear compared to the $a=1$ case.
This shows that SOC model with cellular automaton mechanism can yield lognormal flux distribution by incorporating advection in the energy dissipation processes.

In the light curve of many AGNs, RMS - flux relation are frequently found. Such tendency can be seen also in our simulation with $a=0.1$ (see figure 6) but not clearly appeared in the other calculation with larger values of $a$. This result makes us more convinced that the mechanism proposed here is in the right direction. 

\begin{figure}[t]
\includegraphics[width=60mm, height=80mm, angle=270]{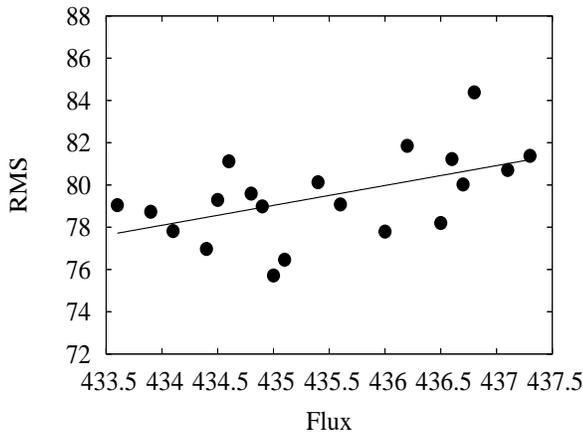}
\caption{RMS - Flux relation derived from the simulated light curve, the solid line is the best fitted straight line to the data points}
\label{fig6}
\end{figure}

\section{Discussion}

The characteristics of simulated light curve which appear in PSD, structure function, flux distribution, and RMS - flux relation are similar with those derived from observation. The slope of the PSD from our simulation is about 1.9, this is in the range of the PSD slope of AGNs \citep[see for example][]{klis95, starling04, uttley02}. 

The structure function can provide some sort of information about the variability time scale. The structure function yielded goes along with the observation \citep[see for example][]{vries03} except the saturation part which provided the hint on longest variability time scale. We can see saturation part in our simulation, but not in the data of 30 years monitoring data of quasars reported by \cite{vries03}. This indicates that the longest variability time scale of quasars is more than 30 years.

The flux distribution of simulated light curve is lognormal which is agree with observation. \cite{Gaskell04} found a lognormal flux distribution in an extreme narrow-line Seyfert 1 Galaxy, and argued that the lognormal intensity distribution of AGN is against the argument that the AGN variations being driven by self organized criticality. That is correct for the old version of SOC model. We, however successfully made some modification of the energy emission mechanism by combining ADAF and cellular automaton mechanism which can yield lognormal distribution, while keeping the other characteristics goes along with the observation. 

This result can revive the SOC model for accretion disk which begin to fade away and more elaborate work based on SOC model is necessary to develop a better model with a hope to finally simulate the real situation in accretion disks. The model developed here is still rough and applied to the accretion disk in a not very detail manner, it's just to show that SOC model can still yield lognormal distribution and RMS-flux relation. 

\acknowledgments

This work is supported by Hibah Peningkatan Kapasitas Institusi ITB Research Fund.
One of the author (CK) got travel support from the Min-OCW Grant, The Netherlands, for research visit to the Kapteyn Astronomical Institute, University of Groningen, The Netherlands in 2010.


\begin{thebibliography}{}
\bibitem[\protect\citeauthoryear{Abramowicz, Czerny, Lasota, \& Szuszkiewicz}{Abramowicz et al.}{1988}]{Abra88} Abramowicz M.A., Czerny B., Lasota J.P., and Szuszkiewicz E., ApJ, 332, 646
\bibitem[\protect\citeauthoryear{Abramowicz, Chen, Kato, Lasota, \& Regev}{Abramowicz et al.}{1995}]{Abra95} Abramowicz M.A., Chen X.M., Kato S., Lasota J.P., and Regev O., ApJ, 438, L37
\bibitem[\protect\citeauthoryear{Abramowicz, Chen, Granath, \& Lasota}{Abramowicz et al.}{1996}]{Abra96} Abramowicz M.A., Chen X.M., Granath M., and Lasota J.P., ApJ, 471, 762
\bibitem[\protect\citeauthoryear{Bak, Tang, \& Wiesenfeld}{Bak et al.}{1988}]{Bak88} Bak P.,, Tang C., and Wiesenfeld K., Phys. Rev. A, 38, 364
\bibitem[\protect\citeauthoryear{de Vries, Becker \& White}{de Vries et al.}{2003}]{vries03} de Vries W.H.,
Becker R.H., White R.L., 2003, ApJ, 126, 1217
\bibitem[\protect\citeauthoryear{Gaskell}{2004}]{Gaskell04} Gaskell C.M., ApJ 612, L21
\bibitem[\protect\citeauthoryear{Gleissner et al.}{2004}]{Gleis04} Gleissner T., Wilms J., Pottschmidt K., Uttley P., Nowak M.A. \& Staubert R., AA 414, 1091
\bibitem[\protect\citeauthoryear{Hawkins}{1996}]{hawkins96} Hawkins M.R.S., MNRAS, 278, 787
\bibitem[\protect\citeauthoryear{Hawkins}{2004}]{hawkins04} Hawkins M.R.S., Baltic Astronomy, 13, 642
\bibitem[\protect\citeauthoryear{Hawkins}{2006}]{hawkins06} Hawkins M.R.S., ASP Conf. Series, ed Gaskell C.M., M$^c$Hardy I.M., Peterson B.M. and Sergeev S.G., 360, 23
\bibitem[\protect\citeauthoryear{Hawkins}{2007}]{hawkins07} Hawkins M.R.S., AA, 462, 581
\bibitem[\protect\citeauthoryear{Ichimaru}{1977}]{Ichimaru77} Ichimaru S., 1977, ApJ, 214, 840
\bibitem[\protect\citeauthoryear{Kato, Fukue \& Mineshige}{Kato et al.}{2008}]{Kato08} Kato S., Fukue J., and Mineshige S., 1998, Black Hole Accretion Disk, Kyoto University Press, Kyoto
\bibitem[\protect\citeauthoryear{Kawaguchi} {2003}]{Kawa03} Kawaguchi T., 2003, ApJ, 593, 69
\bibitem[\protect\citeauthoryear{Kunjaya, Mineshige \& Ivezi\'{c}}{2005}]{Kunjaya05} Kunjaya C., Mineshige S.,and Ivezi\'{c}, 2005, in Proc APRIM, ed Premadi P.W., Hidayat T.,, Mahasena P., ITB Press 
\bibitem[\protect\citeauthoryear{Lehto}{1989}]{Lehto89} Lehto H.J., 1989, in 
Proc 23rd ESLAB Symp on Two Topics in X-Ray Astronomy, ed J. Hunt, B. Battrick, 
Dordrecht ESA SP-296, p499
\bibitem[\protect\citeauthoryear{Lochner, Swank \& Szymkowiak}{Lochner et al.}{1991}]{Lochner91}
Lochner J.C., Swank J.H., and Szymkowiak A.E., 1991, ApJ, 376, 295
\bibitem[\protect\citeauthoryear{Mineshige, Ouchi \& Nishimori}{Mineshige et al.}{1994a}]{mineshige94a}
Mineshige S., Ouchi M.B., and Nishimori H., 1994, PASJ, 46, 97
\bibitem[\protect\citeauthoryear{Mineshige, Takeuchi \& Nishimori}{Mineshige et al.}{1994b}]{mineshige94b}
Mineshige S., Takeuchi M., and Nishimori H., 1994, ApJ, 435, L125
\bibitem[\protect\citeauthoryear{Narayan, \& Yi}{Narayan and Yi}{1994}]{Narayan94}
Narayan R., and Yi I., 1994, ApJ, 428, L13
\bibitem[\protect\citeauthoryear{Narayan, \& Yi}{Narayan and Yi}{1995a}]{Narayan95a}
Narayan R., and Yi I., 1995, ApJ, 444, 231
\bibitem[\protect\citeauthoryear{Narayan, \& Yi}{Narayan and Yi}{1995b}]{Narayan95b}
Narayan R., and Yi I., 1995, ApJ, 452, 710
\bibitem[\protect\citeauthoryear{Narayan, Yi \& Mahadevan}{Narayan et al.}{1995c}]{Narayan95c}
Narayan R., and Yi I., 1995, Nature, 374, 623
\bibitem[\protect\citeauthoryear{Ohshuga, Mineshige, \& Watarai}{Ohshuga et al.}{2003}]{Ohshuga03}
Ohshuga K., Mineshige S., Watarai K., 2003, ApJ, 596, 429
\bibitem[\protect\citeauthoryear{Ohshuga, Mori, Nakamoto, \& Mineshige}{Ohshuga et al.}{2005}]{Ohshuga05}
Ohshuga K., Mori M., Nakamoto T., Mineshige S., 2005, PASJ, 57, 513
\bibitem[\protect\citeauthoryear{Simonetti, Cordes, \& Heeschen}{Simonetti et al.}{1985}]{simonetti85}
Simonetti J.H., Cordes J.M., and Heeschen D.S., 1985, ApJ, 296,46
\bibitem[\protect\citeauthoryear{Starling, Siemiginowska \& Uttley}{Starling et al.}{2004}]{starling04}
Starling R.L.C., Siemiginowska A., and Uttley P., 2004, MNRAS 347, 67
\bibitem[\protect\citeauthoryear{Takeuchi \& Mineshige,}{Takeuchi and Mineshige}{1997}]{Takeuchi97}
Takeuchi M., and Mineshige S., 1997, ApJ, 486, 160
\bibitem[\protect\citeauthoryear{Terrell} {1972}]{Terrell72} Terrel N.J., ApJL 174, L35
\bibitem[\protect\citeauthoryear{Uttley, M$^c$Hardy, \& Papadakis}{Uttley et al.}{2002}]{uttley02}
Uttley P., M$^c$Hardy I.M., and Papadakis I.E., 2002, MNRAS, 332, 231
\bibitem[\protect\citeauthoryear{Uttley, \& M$^c$Hardy}{Uttley and McHardy}{2001}]{uttley01}
Uttley P., M$^c$Hardy I.M., 2001, MNRAS, 323, L26
\bibitem[\protect\citeauthoryear{Uttley, M$^c$Hardy, \& Vaughan}{Uttley et al.}{2005}]{uttley05}
Uttley P., M$^c$Hardy I.M., Vaughan S., 2005, MNRAS 359, 345
\bibitem[\protect\citeauthoryear{Van der Klis}{1995}]{klis95} Van der Klis M., 1995, in
X-ray Binaries, edited by Walter H.G. Lewin, Jan van Paradijs, Edward
P.J. van den Heuvel, Cambridge University Press, p.252
\bibitem[\protect\citeauthoryear{Watarai, Mizuno, \& Mineshige}{Watarai et al.}{2001}]{Watarai01}
Watarai K., Mizuno T., Mineshige S., 2001, ApJ, 549, 77
\end{thebibliography}
\end{document}